\documentclass[3p,12pt]{elsarticle}
\usepackage{graphicx}
\usepackage{amssymb}
\usepackage{amsthm}
\usepackage{amsmath}

\newcommand{\subeqaa}[1]
{
 #1_0 + #1_1 x_1 + #1_2 x_2 + #1_3 x_3
 + #1_4 x_1^2 + #1_5 x_1 x_2
}
\newcommand{\subeqab}[1]
{#1_6 x_1 x_3 + #1_7 x_2^2 + #1_8 x_2 x_3 + #1_9 x_3^2}

\newcommand{\lrs}[1]{\left(#1\right)}
\newcommand{\detb}[4]
{
 \left|
 \begin{array}{ll}
  #1 & #2\\
  #3 & #4
 \end{array}
 \right|
}
\newcommand{\detc}{\detb{c_3}{c_5}{b_0}{b_1}}
\newcommand{\detd}{\detb{b_0}{b_1}{a_0}{a_1}}
\newcommand{\dete}{\detb{b_1}{b_2}{c_5}{c_6}}
\newcommand{\detf}{\detb{a_1}{a_2}{b_1}{b_2}}

\newcommand{\fracb}[1]{\displaystyle\frac{#1}{b_1}}

\newcommand{\eqasub}[3]{$#1$&   $#2$&   $#3$ & $a_3 c_4$}

\newcommand{\eqaaa}{a_3 b_1 c_2}
\newcommand{\eqaab}{a_3 c_1}
\newcommand{\eqaac}{c_3}

\newcommand{\eqaba}{a_3 b_1 c_2 - a_2 b_1 c_3}
\newcommand{\eqabb}{a_2 b_1}

\newcommand{\eqaca}{- a_2 b_1 c_3}
\newcommand{\eqacb}{a_2 b_1 + a_3 c_1}

\newcommand{\eqadb}{- a_1 c_3}
\newcommand{\eqadc}{a_1 + c_3}

\newcommand{\eqaec}{a_1}
\newcommand{\eqafb}{a_2 b_1 - a_1 c_3}
\newcommand{\eqaha}{- a_2 b_1 c_3 - a_3 b_0 c_4}
\newcommand{\eqaja}{a_3 b_1 c_2 - a_2 b_1 c_3 - a_3 b_0 c_4}
\newcommand{\eqakb}{a_3 c_1 - a_1 c_3}
\newcommand{\eqamb}{a_2 b_1 + a_3 c_1 - a_1 c_3}
\newcommand{\eqaoa}{a_3 b_1 c_2 - a_3 b_0 c_4}

\newcommand{\symb}{$\times$}
\newcommand{\Rossler}{R\"{o}ssler\ }

\newtheorem{proposition}{Proposition}

\theoremstyle{definition}
\newtheorem{definition}{Definition}

\journal{Physics Letters A}
\date{}

\nocite*
\begin{document}

\begin{frontmatter}



  \title{Analytic reconstruction of some dynamical systems}


\author {V. Gorodetskyi}

\ead{gorodetskyi@iee.kpi.ua; v.gorodetskyi@gmail.com}


\author {M. Osadchuk}

\address{National Technical University of Ukraine ``Kyiv Polytechnic
  Institute'', 37 Prospect Peremogy, Kyiv 03056, Ukraine}

\begin{abstract}
  We propose a reconstruction of the initial system of ordinary
  differential equations from a single observed variable. The
  suggested approach is applied to a certain class of systems which
  includes, in particular, the \Rossler system and other chaotic
  systems. We develop relations and a method to pass from a model that
  involves the observable and its time derivatives to a real original
  system. To this end, we first find a set of candidates of the system
  in an analytic way. After that, by additionally studying the system, we
  make a choice for the sought system.
\end{abstract}

\begin{keyword}
  ordinary differential equations, observable, reconstruction,
  original system.
\end{keyword}

\end{frontmatter}

\nocite{*}

\section{Introduction}

The problem of reconstruction of a dynamical system from time
series is topical in many fields of human activities. Various
aspects and approaches to this problem have been treated in
numerous studies. Most often, this problem is solved using
numerical methods. For example, the authors in~\cite{CH} have used
a Legendre polynomial approximation of the vector field flow. The
same flow method has been used in~\cite{BH} to reconstruct maps or
differential equations with hidden variables. Alternative
algorithms have been proposed by many authors.
In~\cite{ID,CYMZA,CYZD}, the coefficients of the unknown system
were sought for on the basis of a hierarchical approach. A
two-stage Taylor series approach~\cite{NJC} has been used
in~\cite{JT} to construct an effective algorithm. Also, an
improved algorithm is proposed in~\cite{W}, where the optimization
becomes more effective via a reduction of the initial parameter
region. To achieve this, the author uses a cumulative backward
differentiation formula. In order to improve the calculation
algorithm, a perturbation method is used in~\cite{K-DS} replacing
the traditional Gauss-Newton or Levenberg--Marquardt methods. The
mean square root method is used in~\cite{LLXC} to find the
coefficients in the Lorenz systems~\cite{Lor} and the Chen
systems~\cite{LCZ} for series with noises. H.~Iba~\cite{I} has
improved the mean square root method by using the genetic
programming, which was also used in~\cite{TG,ISSMG}. A
simplification of the calculation algorithm was proposed
in~\cite{D-SF} by reducing the number of the sought system
parameters using a known relation between them. Good
reconstruction results can be obtained via the Bock
algorithm~\cite{B,BSS}. A probability approach to determine the
coefficients of the equations, which generate the time series, has
been used in~\cite{DP}, and for forecasting of the time series
without constructing the model in~\cite{NP}. In~\cite{AFLM}, the
equation coefficients were determined by comparing the type of the
attractor, the first return map, and the largest Lyapunov exponent
that were obtained from the original system and the proposed
model. A synchronization method was used for obtaining a model by
using vector time series in~\cite{BRT}. A similar approach was
used in~\cite{TZDJ,MM}. To forecast the time series, traditional
reconstruction methods are widely used together with relatively
new approaches, including neural networks~\cite{D}, radial basis
functions~\cite{PJM}, fuzzy modeling~\cite{GW}, wavelet
networks~\cite{CHFH}, etc.

The reconstruction problem we will be dealing with consists in a
structure selection for an autonomous differential system with a
polynomial right-hand side from a single observable variable. From
this point of view, the approach proposed
in~\cite{Gousb43,Gousb44,Gousb46,GL} is interesting.
In~\cite{CLA}, the following questions arising in the problem of
global reconstruction were formulated: “(1) Is it possible to
create a model, within a certain class of models, that describes
the initial time series? (2) Is the model unique? If not, what is
the degree of nonuniqueness within the class of models treated?”
Answering these questions in this reference, numerical methods
were important, in particular, the genetic algorithm was used.

In this paper, we will show that the structure of the model can be
obtained for a certain class of systems in terms of mainly strict
mathematical transformations with a minimal use of computational
operations. In addition to the structure, this approach allows
also to obtain relations between the coefficients of the sought
equations, which brings the researcher closer to an ideal solution
of the reconstruction problem, which is to find an exact model for
the original system with a unique true choice for the
coefficients. A procedure to solve the reconstruction problem
could be as follows.
\begin{enumerate}
\item Using a numerical procedure, construct a standard
  system~\cite{Gousb44} using the observable and its derivatives.
\item Using analytic relations between the coefficients of the
  standard and the original systems obtain a family of candidate
  systems ~\cite{Ljung} containing the original system.
\item Using additional information (if this information suffices)
  choose a unique real original system from the family of candidate
  systems.
\item If the additional information does not allow to make the choice,
  conduct an additional investigation. It is assumed that conditions
  of the experiment allow for conducting such an investigation.
\end{enumerate}
It can be seen from the given procedure that it contains only one
typical numerical operation (item 1) that can be repeated if
necessary.

The paper is organized as follows. In Section~\ref{sec:2}, we
consider a relatively simple class of systems that could have
chaotic dynamics. For such a class, we deduce relations that could
be used for reconstructing the unknown system from a single
observable. Section~\ref{sec:3.1} contains examples where we apply
these relations to find analogs of the \Rossler system that would
have the same observable as the original system.
Sections~\ref{sec:3.2},~\ref{sec:3.3} deal with finding, in a
given class, all systems that can be the original system (items~1
and~2 of the above procedure). In Section~\ref{sec:3.4}, we
discuss choosing the original system from the set of candidate
systems (items~3 and~4 from the above procedure).

\section{Essentials of the approach}\label{sec:2}

Consider a system
\begin{equation}
  \label{eq:1}
  \dot x_i = X_i(x_1,\dots, x_i,\dots,x_n),
\end{equation}
where the variables $x_i$, $i=1,\dots,n$, define the state of the
process, $X_i$ are polynomial functions. Following~\cite{Gousb44},
we will call system~(\ref{eq:1}) the Original System~(OS). Assume
we know one solution of~(\ref{eq:1}), a function~$x_i(t)$. Since,
without any further information about system~(\ref{eq:1}), it can
not be recovered from data about only one observable $x_i(t)$,
system~(\ref{eq:1}) is often replaced with a Standard System (SS),
\begin{equation}
  \label{eq:2}
  \left\{
    \begin{aligned}
      \dot y_1&=y_2,\\
      \dot y_2&=y_3,\\
      &\dots\\
      \dot y_n&=Y(y_1,\dots,y_n),
    \end{aligned}
  \right.
\end{equation}
where $y_1(t)\equiv x_i(t)$, $Y(y_1,\dots,y_n)$ is often taken to be a
polynomial or a rational function. Here, many properties of
systems~(\ref{eq:1}) and~(\ref{eq:2}) are the same.

It is clear that, in a general case, such a model can not give an
exact realization of the physics of the process the way the OS does.
Also, the function~$y_1(t)$ obtained when
system~(\ref{eq:2}) is reconstructed could be approximately the same
as the function~$x_i(t)$, $y_1(t)\approx x_i(t)$. At the same time,
there are particular cases where $x_i(t)$ and $y_1(t)$ coincide, that
is, there is a system of type~(\ref{eq:2}) such that $y_1(t)$, being
its solution, satisfies the identity
\begin{equation}
  \label{eq:3}
  y_1(t)\equiv x_i(t).
\end{equation}
Such an exact substitution can be performed if there exists an inverse
standard transformation (IST) that connects the variables
in~(\ref{eq:1}) and~(\ref{eq:2})~\cite{Gousb43}.

It is clear that numerical methods can be applied to a wider class of
systems, however, it may be useful sometimes to use an analytic approach for
finding the OS. The complexity of this approach depends,
among other things, on the general form of the OS,
namely, on the number of the variables and the form of the polynomials
that enter the right-hand side of system~(\ref{eq:1}). For example,
even if the system contains three variables and the degree of the
polynomials is two, the general form of system~(\ref{eq:1}) can be
fairly complex,
\begin{equation}\label{eq:4}
\left\{
 \begin{aligned}
  \dot{x}_1=&\subeqaa{a}\\
  &+\subeqab{a},\\
  \dot{x}_2=&\subeqaa{b}\\
  &+\subeqab{b},\\
  \dot{x}_3=&\subeqaa{c}\\
  &+\subeqab{c}.
 \end{aligned}
\right.
\end{equation}
Correspondingly, if the transformations that connect
systems~(\ref{eq:4}) and~(\ref{eq:2}) exist, they will also be
complex. It was suggested in~\cite{LLS,LLG} to use the ansatz library
that permits to obtain various versions of the OS from
relations between the coefficients of the OS and the SS.

Sometimes, it may happen that a model for complex processes, including
chaotic processes, can have a rather simple form. Although the general
system~(\ref{eq:4}) contains $18$ nonlinear terms, the Lorenz
system~\cite{Lor}, for example, contains only $2$ such terms,
\begin{equation}
  \label{eq:5}
  \left\{
    \begin{aligned}
      \dot x_1&= -\sigma (x_1-x_2),\\
      \dot x_2&= -x_1x_3 - x_2 + \rho x_1,\\
      \dot x_3&=-\beta x_3 + x_1 x_2,
    \end{aligned}
  \right.
\end{equation}
where $\sigma,\rho,\beta$ are constant parameters. The \Rossler system~\cite{Ross},
\begin{equation}
  \label{eq:6}
  \left\{
    \begin{aligned}
      \dot x_1&= -x_2-x_3,\\
      \dot x_2&= x_1+a x_2,\\
      \dot x_3&=b-cx_3 + x_1x_3,
    \end{aligned}
  \right.
\end{equation}
where $a,b,c$ are parameters, has even simpler
form. System~(\ref{eq:6}) has only one nonlinear term which is a
product of two different variables. As opposed to the general form of
system~(\ref{eq:4}), systems~(\ref{eq:5}) and~(\ref{eq:6}) do not
contain squares of the variables. Having this in mind, one can
consider replacing the general type system~(\ref{eq:4}) with a
simplified system of the form
\begin{equation}
  \label{eq:7}
  \left\{
    \begin{aligned}
      \dot x_1&= a_0 + a_1 x_1 + a_2 x_2 + a_3 x_3,\\
      \dot x_2&= b_0 + b_1x_1 + b_2 x_2 + b_3 x_3,\\
      \dot x_3&=c_0 + c_1 x_1 + c_2 x_2 + c_3 x_3 +
      c_4 x_1x_2 + c_5 x_1 x_3 + c_6 x_2 x_3.
    \end{aligned}
  \right.
\end{equation}

\begin{definition}\label{D:1}
  We say that a system belongs to the \Rossler class (R-class) if it is
  of the form~(\ref{eq:7}) and only one of the
  coefficients~$c_4,c_5,c_6$ is distinct from zero.
\end{definition}

Choose the variable $x_2$ to be the observable in the \Rossler system,
that is, $y_1=x_2$. Its reconstruction, by~\cite{GL}, can be obtained
in the form
\begin{equation}
  \label{eq:8}
  \left\{
    \begin{aligned}
      \dot y_1=& y_2,\\
      \dot y_2=& y_3\\
      \dot y_3=&A_0 + A_1 y_1 + A_2 y_2 + A_3 y_3 + A_4 y_1^2
      + A_5 y_1 y_2 + A_6 y_1 y_3 \\
      & + A_7 y_2^2 + A_8 y_2 y_3 +
      A_9 y_3^2,
    \end{aligned}
  \right.
\end{equation}
where $A_0,\dots, A_9$ are constants with
\begin{equation}
  \label{eq:9}
  A_9=0.
\end{equation}
It is easy to show that relation~(\ref{eq:9}) holds not only for the
\Rossler system but for systems of a more general type.

\begin{definition}\label{D:2}
  We say that a system belongs to a shortened \Rossler class
  ($SR$-class), if it is of the $R$-class and $b_3=0$, that is, a
  general $SR$-system is of the form
\begin{equation}
  \label{eq:10}
  \left\{
    \begin{aligned}
      \dot x_1&= a_0 + a_1 x_1 + a_2 x_2 + a_3 x_3,\\
      \dot x_2&= b_0 + b_1x_1 + b_2 x_2,\\
      \dot x_3&=c_0 + c_1 x_1 + c_2 x_2 + c_3 x_3 +
      c_4 x_1x_2 + c_5 x_1 x_3 + c_6 x_2 x_3.
    \end{aligned}
  \right.
\end{equation}
\end{definition}

\begin{proposition}\label{P:1}
  If the OS has form~(\ref{eq:10}), then its
  reconstruction to a SS will have the
  form
\begin{equation}\label{eq:11}
\left\{
 \begin{aligned}
  \dot{y}_1=&y_2,\\
  \dot{y}_2=&y_3,\\
  \dot{y}_3=&A_0 + A_1 y_1 + A_2 y_2 + A_3 y_3 + A_4 y_1^2 +
  A_5 y_1 y_2 + A_6 y_1 y_3\\
  &+A_7 y_2^2 + A_8 y_2 y_3
 \end{aligned}
\right.
\end{equation}
and the relations between the coefficients in systems~(\ref{eq:10})
and~(\ref{eq:11}) is given by
\begin{equation}\label{eq:12}
\left\{
 \begin{aligned}
  A_0=&a_3\detb{c_0}{c_1}{b_0}{b_1}+\fracb{1}\detc\detd,\\
  A_1=&a_3\lrs{\detb{b_1}{b_2}{c_1}{c_2}-b_0 c_4}\\
  &+\fracb{1}\lrs{\detc\detf+\dete\detd},\\
  A_2=&\detb{b_1}{b_2}{a_1}{a_2}+a_3 c_1+\fracb{c_5}\detd+
  \fracb{a_1+b_2}\detb{b_0}{b_1}{c_3}{c_5},\\
  A_3=&a_1+b_2+\fracb{1}\detc,\\
  A_4=&\fracb{1}\dete\detf-a_3 b_2 c_4,\\
  A_5=&a_3 c_4+\fracb{c_5}\detf+\fracb{a_1+b_2}\detb{c_5}{c_6}{b_1}{b_2},\\
  A_6=&\fracb{1}\detb{b_1}{b_2}{c_5}{c_6},\\
  A_7=&-\fracb{c_5}\lrs{a_1+b_2},\\
  A_8=&\fracb{c_5},\\
  b_1\not=&0.
 \end{aligned}
\right.
\end{equation}
\end{proposition}

A proof of Proposition~\ref{P:1} is given in \ref{app:1}.

Hence, one can infer that if the reconstruction of the unknown
OS from one observable is of form~(\ref{eq:11}), then the
OS can be of the SR-class. This fact can be used in the
analytical search for the form of the OS, replacing the
calculation procedure.

The \Rossler system~(\ref{eq:6}), with the notations as in~(\ref{eq:10}),
can be represented as
\begin{equation}\label{eq:13}
  \left\{
    \begin{aligned}
      \dot{x}_1&=a_2 x_2 + a_3 x_3,\\
      \dot{x}_2&=b_1 x_1+ b_2 x_2,\\
      \dot{x}_3&=c_0 +  c_3 x_3 + c_5 x_1 x_3,
    \end{aligned}
  \right.
\end{equation}
where $a_2=-1$, $a_3=-1$, $ b_1=1$, $b_2=a$, $c_0=b$, $c_3=-c$,
$c_5=1$. Since $a_0 = a_1 = b_0 = c_1 = c_2 = c_4 = c_6 = 0$ in this
case, the expressions in~(\ref{eq:12}) for the \Rossler system become
\begin{equation}
  \label{eq:14}
  \left\{
  \begin{array}{ll}
    A_0 = a_3 b_1 c_0 = -b,\quad&
    A_1 = -a_2 b_1 c_3 = -c,\\
    A_2 = a_2b_1 - b_2 c_3 = -1 + ac ,\quad&
    A_3 = b_2 + c_3 = a - c ,\\
    A_4 = a_2b_2 c_5 = - a , \quad&
    A_5 = c_5 ( b_2^2 / b_1 - a_2 ) = a^2 + 1,\\
    A_6 = A_7 = -b_2 c_5 / b_1 = - a ,\quad&
    A_8 = c_5 / b_1 = 1.
  \end{array}
  \right.
\end{equation}
Using relations~(\ref{eq:14}) one can obtain systems of
type~(\ref{eq:11}) that would have the solution~$y_1(t)$
coinciding with a solution~$x_2(t)$ of the original \Rossler system.

Let us state an obvious general result that we will use in the following.

\begin{proposition}
  \label{P:2}
  If two different OS's have the same solution considered
  as the observable $y_1(t)$, then the SS reconstructed
  from this observable will be the same for both OS's.
\end{proposition}

\section{Results}\label{sec:3}
\subsection{Analogs of the \Rossler system}\label{sec:3.1}

Using Proposition~\ref{P:2} and relations~(\ref{eq:14}) one can obtain
different versions of the OS, which would have the same
structure, that is, the nonzero coefficients in the right-hand sides
will be the same as in the \Rossler system. The values of the
coefficients of the new system will differ from those in the \Rossler
system but the function $x_2(t)$ will coincide with the function
$x_2(t)$ in the \Rossler system.

\begin{figure}[htb]
\begin{center}
  \includegraphics[width=12cm]{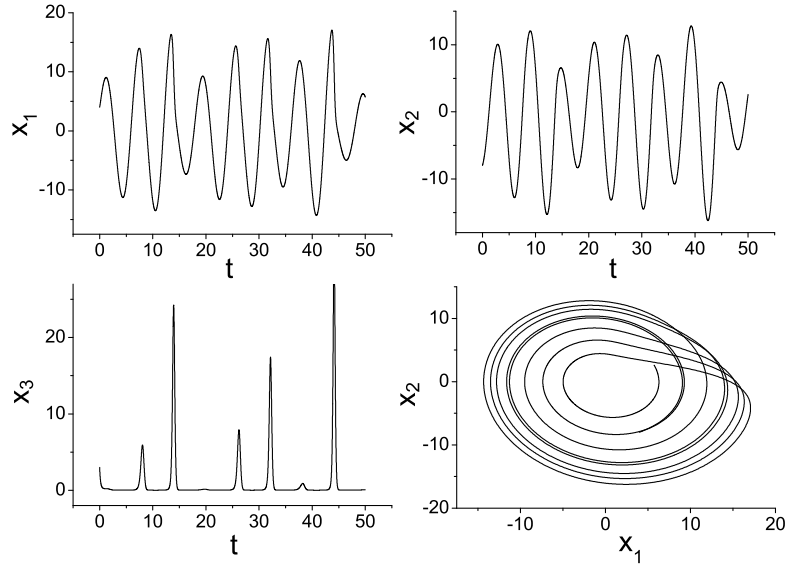}
\end{center}
  \caption{The functions $x_1(t)$, $x_2(t)$, $x_3(t)$, $x_2(x_1)$ in
  the original \Rossler system~(\ref{eq:13}). Numeric integration is
  carried out with the fourth-order Runge-Kutta method on the time
  interval of 50 sec. with step 0.005 sec.}\label{fig:1}
\end{figure}
\begin{figure}[htb]
\begin{center}
  \includegraphics[width=12cm]{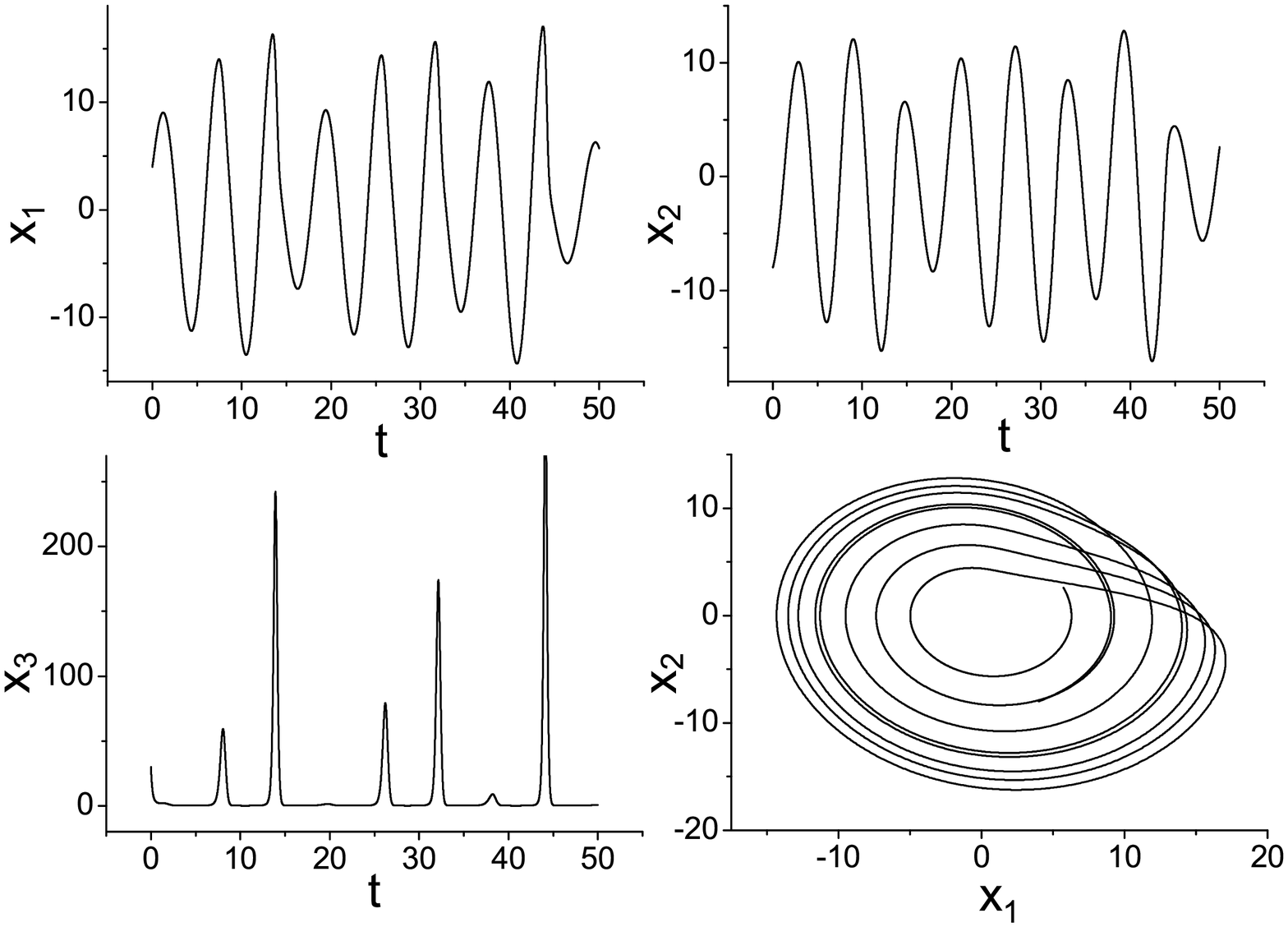}
\end{center}
  \caption{The same as in Fig.~\ref{fig:1} for the \Rossler system with
    the values of $c_0$ and $a_3$ changed.}\label{fig:2}
\end{figure}

As an example, consider a reconstruction of the \Rossler system
for $a = 0.15$, $b = 0.2$, $c = 10$. The functions $x_1(t)$, $x_2(t)$,
$x_3(t)$, $x_2(x_1)$ are presented for this case in
Fig.~\ref{fig:1}. For system~(\ref{eq:13}), we get $a_2 = a_3 =
-1, b_1 = 1, b_2 = 0.15, c_0 = 0.2, c_3 = -10, c_5 = 1$.
By~(\ref{eq:14}), $A_0 = a_3b_1c_0$. If, for example, $b_1$ is
unchanged, and the coefficient $c_0$ is increased by $10$ times, then
for $A_0$ to be unchanged, the absolute value of $a_3$ must be
reduced also by $10$ times. New values of the coefficients will be
$c_0 = 2, a_3 = -0.1 $. Since these changes do not change the
value of $A_0$ and the coefficients $A_1,\dots. A_8$ do not depend
on $c_0$ and $a_3$, hence remain unchanged, Proposition~\ref{P:2}
can be applied. Hence, the solution $x_2(t)$ of the new OS
must coincide with $x_2(t)$ of the original \Rossler
system. This is illustrated in Fig.~\ref{fig:2}.
The graphs show that the functions $x_1(t)$ and $x_2(t)$ remain
unchanged, see Fig.~\ref{fig:1}, and, for the function $x_3$, the
scale has changed. The scaling problem in more detail has been
considered in~\cite{CLA}.

\begin{figure}[htb]
\begin{center}
  \includegraphics[width=12cm]{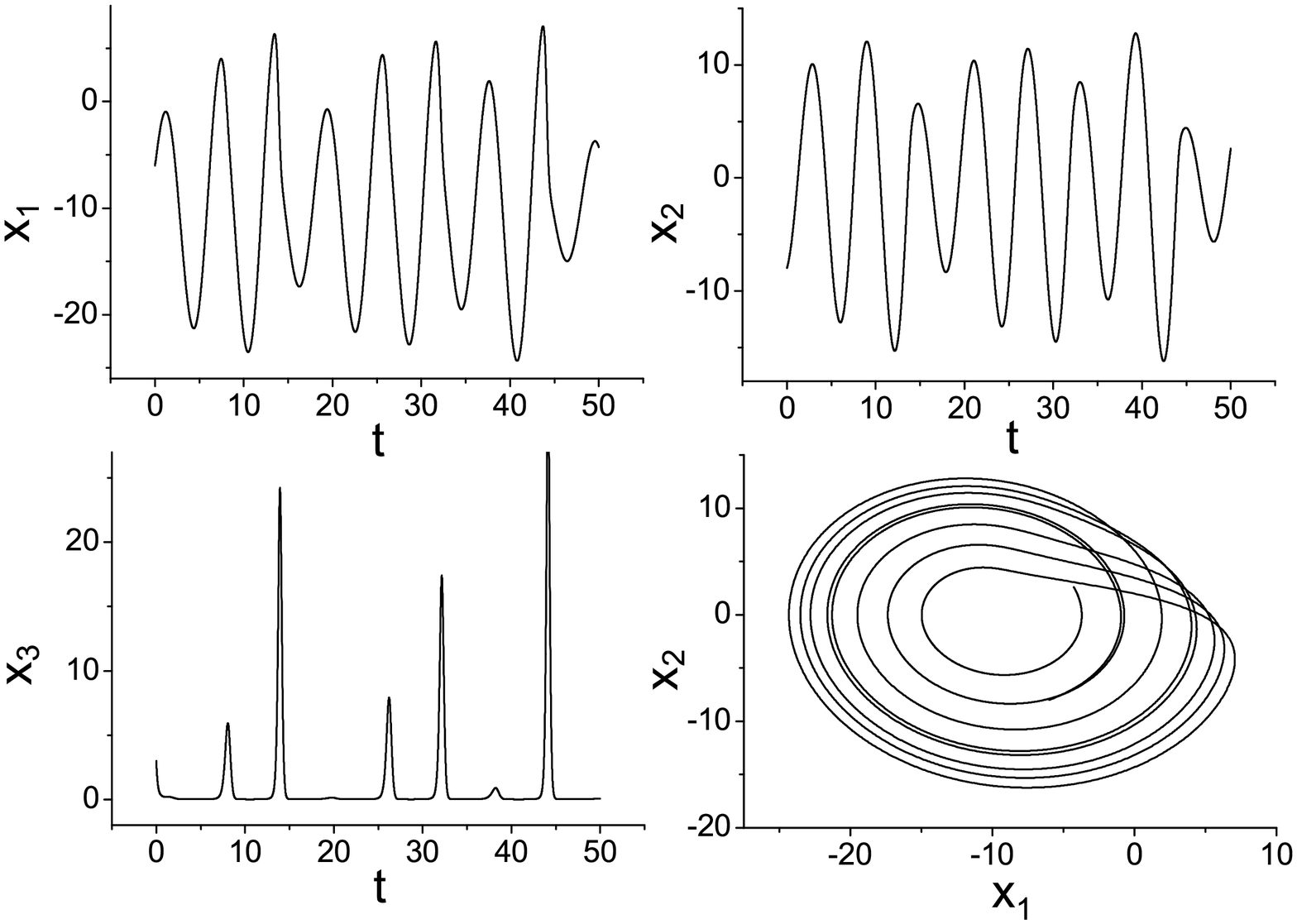}
\end{center}
  \caption{The same as in Fig.~\ref{fig:1} for
    system~(\ref{eq:15}).}\label{fig:3}
\end{figure}

Using relations~(\ref{eq:12}) it is also possible to obtain a
modification of the \Rossler system with a change of the structure, for
example, like the following:
\begin{equation}\label{eq:15}
\left\{
\begin{aligned}
  \dot{x}_1&=a_2 x_2 + a_3 x_3,\\
  \dot{x}_2&=b_0+ b_1 x_1+b_2x_2,\\
  \dot{x}_3&=c_0+ c_5 x_1 x_3.
\end{aligned}
\right.
\end{equation}
Comparing systems~(\ref{eq:15}) and~(\ref{eq:13}), one can see
that in system~(\ref{eq:15}) $c_3=0$ but $b_0\neq 0$. Then, for
system~(\ref{eq:15}), we get from~(\ref{eq:12}) that
\begin{equation}
  \label{eq:16}
  \left\{
  \begin{array}{ll}
    A_0 = a_3 b_1 c_0 ,\quad&
    A_1 = a_2 b_0 c_5 ,\\
    A_2 = a_2 b_1 + b_0 b_2 c_5 / b_1,\quad&
    A_3 = b_2 - b_0 c_5 / b_1 ,\\[1.5mm]
    A_4 = a_2 b_2 c_5 ,\quad&
    A_5 = c_5 ( b_2^2 / b_1 - a_2 ) ,\\
    A_6 = A_7 = -b_2 c_5 / b_1 ,\quad&
    A_8 = c_5 / b_1.
  \end{array}
  \right.
\end{equation}
The analysis given in \ref{app:2} shows that by setting
$b_0=c$ and leaving $a_2 $, $a_3 $, $b_1 $, $b_2 $, $c_0 $, $c_5 $ as
in system~(\ref{eq:13}), the values of the coefficients in the
reconstruction~(\ref{eq:16}) will be the same as the coefficients in
the reconstruction~(\ref{eq:14}). Consequently, system~(\ref{eq:15})
satisfies the conditions in Proposition~\ref{P:2}. Hence the solution
$x_2(t)$ of system~(\ref{eq:15}) will coincide with the solution
$x_2(t)$ of system~(\ref{eq:13}) for an appropriate choice of the initial
conditions. This is illustrated in Fig.~\ref{fig:3}, where one can
also see that the function $x_3(t)$ coincides with that for the
\Rossler system, and $x_1(t)$ is shifted as compared with $x_1(t)$ for
system~(\ref{eq:13}).

\subsection{Finding the OS from its reconstruction in the
  form of a SS}\label{sec:3.2}

\begin{figure}[ht]
\begin{center}
  \includegraphics[width=12cm]{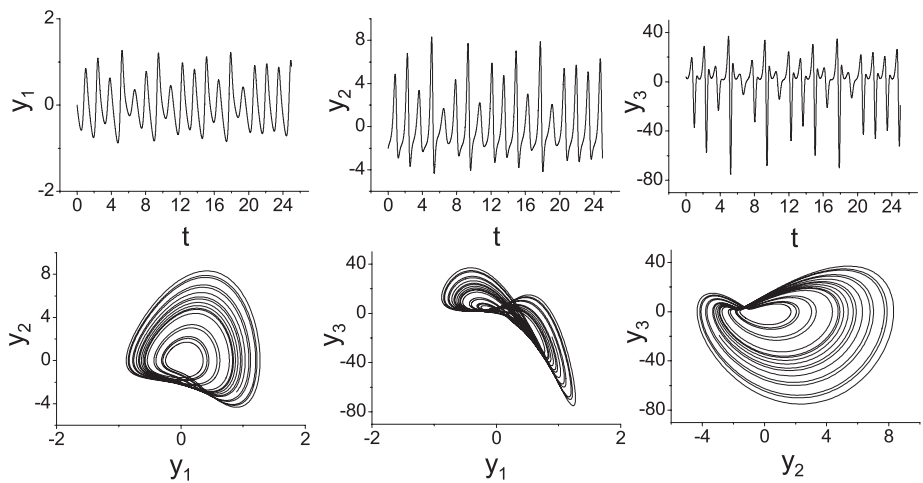}
\end{center}
  \caption{The functions and phase portraits for system~(\ref{eq:SS4});
    $y_1(t)$ is the observable $x_2(t)$ in
    system~(\ref{eq:OS6}).}\label{fig:4}
\end{figure}

In some cases, the proposed approach could be effective when
recovering an unknown model from its reconstruction~(\ref{eq:11}). As
an example, consider the system
\begin{equation}\label{eq:OS6}
\left\{
 \begin{aligned}
  \dot{x}_1&=a_2 x_2 + a_3 x_3,\\
  \dot{x}_2&=b_1 x_1,\\
  \dot{x}_3&=c_1 x_1 + c_3 x_3 + c_4 x_1 x_2,
\end{aligned}
\right.
\end{equation}
where $a_2 = -25.4, a_3 = 1, b_1 = 1, c_1 = 15.4, c_3 = -10,
c_4 = -143 $. This and all subsequent systems were solved by
applying the fourth-order Runge-Kutta method on the time interval
of 25 sec. with step 0.001 sec. The obtained time series for
$x_2(t)$ in system~(\ref{eq:OS6}) was taken to be a unique
observable, that is, we took $y_1(t)\equiv x_2(t)$ for the
reconstruction in the form of the SS. The
reconstructed SS of the form~(\ref{eq:11}) has the
coefficients shown in Table~\ref{Tbl:1}. The reconstruction was
obtained by applying both a numerical method, similar
to~\cite{Gousb43}, and an analytical method using
relations~(\ref{eq:12}). As Table~\ref{Tbl:1} shows, the errors in
the values of the coefficients in the reconstruction obtained
numerically are small. In the following, when making necessary
transformations we used exact values obtained analytically and
shown in the right column of the table.
\begin{table}
  \centering
  \begin{tabular}{|c|c|c|}
    \hline
    Coefficients of SS & Numerical method & Analytical method\\
    \hline
    $A_0$ &     $-3.1427\cdot 10^{-5}$ &    $0$\\
    \hline
    $A_1$ &     $-253.9956$     &   $-254$\\
    \hline
    $A_2$ &     $-9.9995$ &         $-10$\\
    \hline
    $A_3$ &     $-9.9998$ &         $-10$\\
    \hline
    $A_4$ &     $6.6653\cdot 10^{-5}$ &     $0$\\
    \hline
    $A_5$ &     $-142.9986$ &           $-143$\\
    \hline
    $A_6$ &     $-2.3495\cdot 10^{-4}$ &    $0$\\
    \hline
    $A_7$ &     $1.2581\cdot 10^{-4}$ &     $0$\\
    \hline
    $A_8$ &     $-2.4444\cdot 10^{-4}$ &    $0$\\
    \hline
    $A_9$ &     $-1.1524\cdot 10^{-5}$ &    $0$\\
    \hline
  \end{tabular}
  \bigskip
  \caption{Values of the coefficients in reconstruction~\eqref{eq:11}
    for the solution~$x_2(t)$ of the OS~\eqref{eq:OS6}
    obtained numerically and analytically using relation~\eqref{eq:12}.
    The value of the coefficient $A_9$ obtained by numerical calculations
    verifies relation~(\ref{eq:9}).}
  \label{Tbl:1}
\end{table}
As a result, the SS has the form
\begin{equation}\label{eq:SS4}
\left\{
 \begin{aligned}
  \dot{y}_1&=y_2,\\
  \dot{y}_2&=y_3,\\
  \dot{y}_3&=A_1 y_1 + A_2 y_2 + A_3 y_3 + A_5 y_1 y_2,
\end{aligned}
\right.
\end{equation}
where $A_1 = -254, A_2 = -10, A_3 = -10, A_5 = -143 $.
Solutions of system~(\ref{eq:SS4}) and its phase portraits are
shown in Fig.~\ref{fig:4}.

We will consider the OS~(\ref{eq:OS6}) to be unknown. To
obtain information about it, we use~(\ref{eq:12})
and~(\ref{eq:SS4}). The results of the analysis given in
\ref{app:3} show that the sought OS has $a_0 =
b_2 = c_0 = c_5 = c_6 = 0 $ and the nonzero coefficients could only be
the following: $a_1 , a_2 , a_3 , b_0 , b_1 , c_1 , c_2 , c_3 , c_4
$. This significantly simplifies system~(\ref{eq:12}) giving
\begin{equation}
  \label{eq:A012345}
  \left\{
  \begin{aligned}
     A_0& = a_1b_0 c_3 - a_3b_0 c_1 ,\\
     A _1&= a_3 b_1 c_2 - a_2 b_1 c_3 - a_3 b_0 c_4 ,\\
     A_2 &= a_2b_1 + a_3c_1 - a_1c_3 ,\\
     A_3 &= a_1 + c_3 ,\\
     A_5 &= a_3c_4 ,\\
     A_4 &= A_6 = A_7 = A_8 = 0.
   \end{aligned}
   \right.
 \end{equation}

 To pass to the reconstruction of the OS, it is necessary
 to determine how many of the 9 nonzero coefficients will be contained
 in the sought equations and which of them. As a simplest version of
 the OS, one can take the following system that has the
 same structure as system~(\ref{eq:SS4}):
\begin{equation}\label{eq:b0a1a2}
\left\{
 \begin{aligned}
  \dot{x}_1&=a_3 x_3,\\
  \dot{x}_2&=b_1 x_1,\\
  \dot{x}_3&=c_1 x_1 + c_2 x_2 + c_3 x_3 + c_4 x_1 x_2.
\end{aligned}
\right.
\end{equation}
The quantities in systems~(\ref{eq:SS4}) and~(\ref{eq:b0a1a2}) are
connected with the relation $x_2 = y_1 $, $a_3 = 1$, $b_1 = 1$, $c_1 =
A_2$, $c_2 = A_1$, $c_3 = A_3 $, $c_4 = A_5 $.

It is clear that system~(\ref{eq:b0a1a2}) is not the only possible one
for the OS recovered from the SS~(\ref{eq:SS4}). Denote by $K$ the
total number of nonzero
coefficients in the right-hand sides of the OS.  We
assume, similarly to~(\ref{eq:b0a1a2}), that $K=6$ for other unknown
OS's (the case where $K\neq 6$ will be considered in
Section~\ref{sec:3.3}). Since relations~(\ref{eq:A012345}) involve $9$
coefficients, when considering candidate systems~\cite{Ljung} to
choose a unique real OS from, it is necessary to check
relations~(\ref{eq:A012345}) for possible combinations of $6$
coefficients from $9$ as to satisfy conditions in
Proposition~\ref{P:2}. To this end, it is sufficient to alternatively
set $3$ out of $9$ coefficients to zero and check whether the
structure of the SS is that of~(\ref{eq:SS4}). Such
transformations are carried out in \ref{app:3}. As the result we get
the OS's $S_1$ -- $S_8$, see Table~\ref{Tbl:2}.

\begin{table}[htb]
  \centering
  \begin{tabular}{|c|c|c|c|c|c|c|c|c|c|c|}
    \hline
    $K$ &   OS  &  \multicolumn{9}{|c|}{Coefficients of OS}\\
    \cline{3-11}
    &       &   $a_1$&  $a_2$&  $a_3$&  $b_0$&  $b_1$&  $c_1$&  $c_2$&  $c_3$&  $c_4$\\
    \hline

    &$S_1$  &       &   &   \symb&  &   \symb&  \symb&  \symb&  \symb&  \symb\\
    \cline{2-11}
    &$S_2$  &       &   \symb&  \symb&  &   \symb&  &   \symb&  \symb&  \symb\\
    \cline{2-11}
    &$S_3$  &       &   \symb&  \symb&  &   \symb&  \symb&  &   \symb&  \symb\\
    \cline{2-11}
    6&  $S_4$&      \symb&  &   \symb&  &   \symb&  &   \symb&  \symb&  \symb\\
    \cline{2-11}
    &$S_5$&         \symb&  &   \symb&  &   \symb&  \symb&  \symb&  &   \symb\\
    \cline{2-11}
    &$S_6$  &       \symb&  \symb&  \symb&  &   \symb&  &   &   \symb&  \symb\\
    \cline{2-11}
    &$S_7$  &       \symb&  \symb&  \symb&  &   \symb&  &   \symb&  &   \symb\\
    \cline{2-11}
    &$S_8$  &       &   \symb&  \symb&  \symb&  \symb&  &   &   \symb&  \symb\\
    \hline

    &$S_9$  &       &   \symb&  \symb&  &   \symb&  \symb&  \symb&  \symb&  \symb\\
    \cline{2-11}
    &$S_{10}$&      &   \symb&  \symb&  \symb&  \symb&  &   \symb&  \symb&  \symb\\
    \cline{2-11}
    &$S_{11}$&      \symb&  &   \symb&  &   \symb&  \symb&  \symb&  \symb&  \symb\\
    \cline{2-11}
    7&$S_{12}$&     \symb&  \symb&  \symb&  &   \symb&  &   \symb&  \symb&  \symb\\
    \cline{2-11}
    &$S_{13}$&      \symb&  \symb&  \symb&  &   \symb&  \symb&  &   \symb&  \symb\\
    \cline{2-11}
    &$S_{14}$&      \symb&  \symb&  \symb&  &   \symb&  \symb&  \symb&  &   \symb\\
    \cline{2-11}
    &$S_{15}$&      \symb&  \symb&  \symb&  \symb&  \symb&  &   \symb&  &   \symb\\
    \hline

    8&$S_{16}$&     \symb&  \symb&  \symb&  &   \symb&  \symb&  \symb&  \symb&  \symb\\
    \hline

    9&$S_{17}$&     \symb&  \symb&  \symb&  \symb&  \symb&  \symb&  \symb&  \symb&  \symb\\
    \hline
  \end{tabular}
  \bigskip
  \caption{OS's corresponding to reconstruction~\eqref{eq:SS4}:
    $K$ is the number of the nonzero coefficients in the right-hand sides of the
    OS's. System~\eqref{eq:OS6} is denoted by~$S_3$,
    system~\eqref{eq:b0a1a2} by~$S_1$, and system~\eqref{eq:k9} by~$S_{17}$.}
  \label{Tbl:2}
\end{table}

\begin{figure}[htp]
\begin{center}
  \includegraphics[width=10cm]{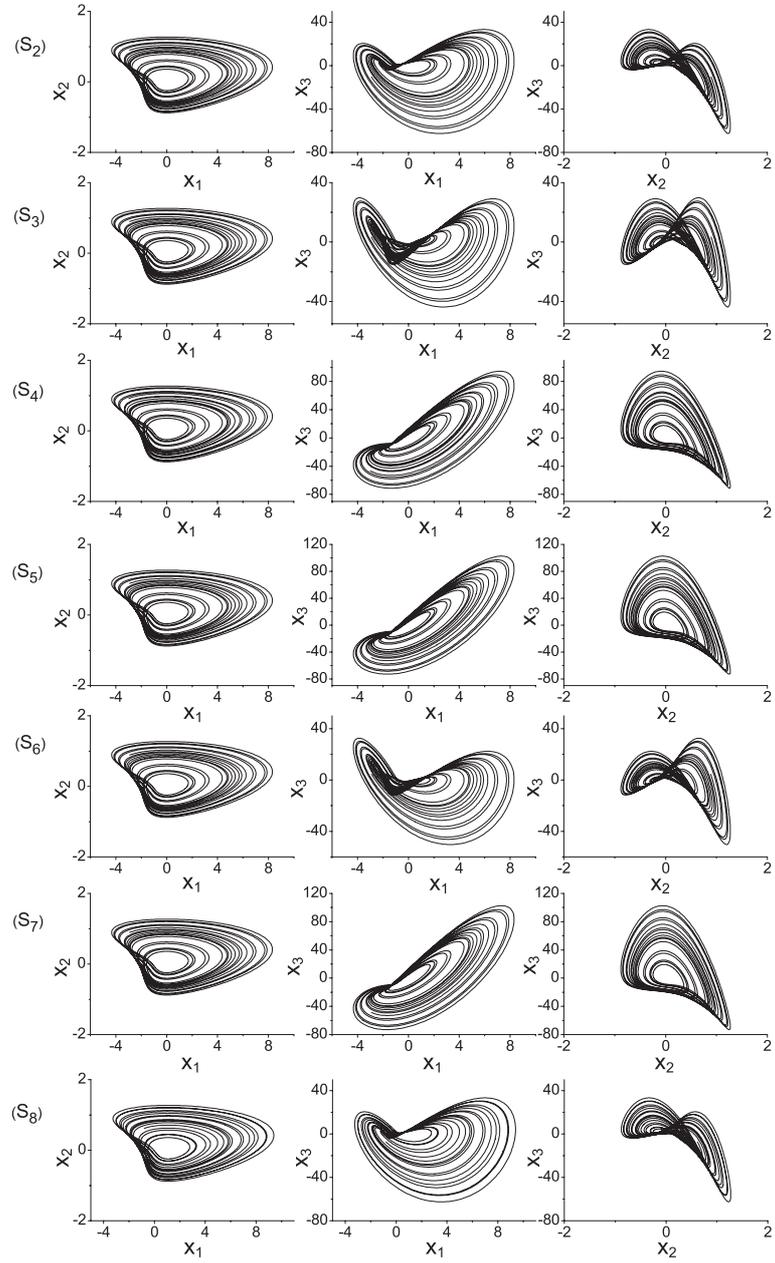}
\end{center}
  \caption{Phase portraits of systems $S_2$ --
    $S_8$.}\label{fig:5}
\end{figure}

The functions $x_2(t)$ for all systems $S_1$ -- $S_8$ are practically
coincide with the time series for $y_1(t)$ used in the reconstruction,
see Fig.~\ref{fig:4}.
Insignificant differences are due to numerical integration errors
and rounding the fractional values of the coefficients obtained
from~(\ref{eq:A012345}) for some OS's. The phase
curves for systems $S_2$ -- $S_8$ are shown in Fig.~\ref{fig:5}.
The phase trajectories of system~$S_1$ are similar to the
trajectories of system~(\ref{eq:SS4}). Relations for the
coefficients for the SS and all OS's
obtained from~(\ref{eq:A012345}) are shown in Table~\ref{Tbl:3}.

\begin{table}[htb]
  \centering
  \begin{tabular}{|c|c|c|c|c|c|}
    \hline
    $K$&    OS & \multicolumn{4}{|c|}{Relations between the coefficients of SS and OS}\\
    \cline{3-6}
    &       &   $A_1$&  $A_2$&  $A_3$&  $A_5$\\
    \hline

    &$S_1$  &       \eqasub{\eqaaa}{\eqaab}{\eqaac}\\
    \cline{2-6}
    &$S_2$  &       \eqasub{\eqaba}{\eqabb}{\eqaac}\\
    \cline{2-6}
    &$S_3$  &       \eqasub{\eqaca}{\eqacb}{\eqaac}\\
    \cline{2-6}
    6&  $S_4$&      \eqasub{\eqaaa}{\eqadb}{\eqadc}\\
    \cline{2-6}
    &$S_5$&         \eqasub{\eqaaa}{\eqaab}{\eqaec}\\
    \cline{2-6}
    &$S_6$  &       \eqasub{\eqaca}{\eqafb}{\eqadc}\\
    \cline{2-6}
    &$S_7$  &       \eqasub{\eqaaa}{\eqabb}{\eqaec}\\
    \cline{2-6}
    &$S_8$  &       \eqasub{\eqaha}{\eqabb}{\eqaac}\\
    \hline

    &$S_9$  &       \eqasub{\eqaba}{\eqacb}{\eqaac}\\
    \cline{2-6}
    &$S_{10}$&      \eqasub{\eqaja}{\eqabb}{\eqaac}\\
    \cline{2-6}
    &$S_{11}$&      \eqasub{\eqaaa}{\eqakb}{\eqadc}\\
    \cline{2-6}
    7&$S_{12}$&     \eqasub{\eqaba}{\eqafb}{\eqadc}\\
    \cline{2-6}
    &$S_{13}$&      \eqasub{\eqaca}{\eqamb}{\eqadc}\\
    \cline{2-6}
    &$S_{14}$&      \eqasub{\eqaaa}{\eqacb}{\eqaec}\\
    \cline{2-6}
    &$S_{15}$&      \eqasub{\eqaoa}{\eqabb}{\eqaec}\\
    \hline

    8&$S_{16}$&     \eqasub{\eqaba}{\eqamb}{\eqadc}\\
    \hline

    9&$S_{17}$&     \eqasub{\eqaja}{\eqamb}{\eqadc}\\
    \hline
  \end{tabular}
  \bigskip
  \caption{Relations between the coefficients of the SS~\eqref{eq:SS4}
  and OS's corresponding to~\eqref{eq:A012345}.}
  \label{Tbl:3}
\end{table}

\subsection{Determining a complete family of candidate
  systems}\label{sec:3.3}

In the previous section, we have considered $8$ possible
OS's that have $6$ coefficients in the right-hand sides. Using
relation~(\ref{eq:A012345}) one can  find the number $K$ of
possible nonzero coefficients in the OS that has
system~(\ref{eq:SS4}) as a SS. First, let us find a lower
bound for $K$, denoted by~$K_\mathrm{min}$.

\begin{definition}
  \label{D:3}
  An OS will be called minimal for the
  SS~(\ref{eq:SS4}) if it permits a reconstruction in the form
  of~(\ref{eq:SS4}) and has $K_\mathrm{min}$ nonzero coefficients, and
  any system of the form~(\ref{eq:10}) with $K<K_\mathrm{min}$ can not
  give a reconstruction in the form of~(\ref{eq:SS4}).
\end{definition}

Let us check whether systems $S_1$ -- $S_8$ are minimal, that is,
whether there exists an OS with $K<6$. To do this, it
is necessary to equate various coefficients in these systems to
zero, but the coefficients $A_1 , A_2 , A_3 , A_5 $ in the
SS should be distinct from zero. It turned out that
there is a unique system with $K=5$ that satisfies this condition,
\begin{equation}\label{eq:k5}
\left\{
 \begin{aligned}
  \dot{x}_1&= a_2 x_2 + a_3 x_3,\\
  \dot{x}_2&= b_1 x_1,\\
  \dot{x}_3&= c_3 x_3 + c_4 x_1 x_2.
\end{aligned}
\right.
\end{equation}
Here, by~(\ref{eq:A012345}),
\begin{equation}
  \label{eq:A1235}
  A_1 = - a_2b_1c_3 ,\quad
  A_2 = a_2b_1 ,\quad
  A_3 = c_3 ,\quad
  A_5 = a_3c_4.
\end{equation}
Although $A_1 , A_2 , A_3 , A_5 $ are nonzero, it follows
from~(\ref{eq:A1235}) that
\begin{equation}
  \label{eq:123}
  A_1=-A_2 A_3.
\end{equation}
By substituting the values of $A_1 , A_2 , A_3 $ from Table
\ref{Tbl:1} into~(\ref{eq:123}), we see that~(\ref{eq:123}) is not
true. Consequently, relations~(\ref{eq:A1235}) do not permit to
obtain a reconstruction of~(\ref{eq:SS4}), and
system~(\ref{eq:k5}) can not be an OS. So, the minimal OS's could
only be those with $K=6$ in Table~\ref{Tbl:2}.

The analysis in~\ref{app:3} shows that the maximal value of
$K$ so that the corresponding OS is reconstructed
from~(\ref{eq:SS4}) is $K_\mathrm{max}=9$. That is, a maximal
OS will contain all the coefficients that enter the right-hand
sides of relations~(\ref{eq:A012345}). It has the form
\begin{equation}\label{eq:k9}
\left\{
 \begin{aligned}
  \dot{x}_1&=a_1 x_1 + a_2 x_2 + a_3 x_3,\\
  \dot{x}_2&=b_0 + b_1 x_1,\\
  \dot{x}_3&=c_1 x_1 + c_2 x_2 + c_3 x_3 + c_4 x_1 x_2.
\end{aligned}
\right.
\end{equation}
Since $K=9$ for this system and the number of known coefficients
in the SS~(\ref{eq:SS4}) is $4$, some values for the
coefficients $a_1,\dots,c_4$ were taken arbitrarily ($a_3=b_1=1$)
whereas the others were determined from~(\ref{eq:A012345}). As the
result we get $a_1 = -5$, $ a_2 = -10 $, $ b_0 = -1$, $
c_1 = 25$, $ c_2 = -61$, $ c_3 = -5$, $ c_4 = -143 $.

Now, to find all possible OS's other than with
$K_\mathrm{min}=6$ and $K_\mathrm{max}=9$, consider possible
OS's with $K=7$ and $K=8$. We can previously estimate
the possible number of OS's for each value of $K$.
Three of the coefficients, $a_3$, $b_1$, and $c_4$, are always
nonzero (see~\ref{app:3}), whereas the other $6$ may or may not be
zero. If $K=8$, the system must have $5$ more nonzero
coefficients. Various cases can be obtained by in turn equating
one of the six unknown coefficients to zero. The number of such
combinations will be $n_8=\binom{6}{1}=6$. Similarly, if $K=7$, we
get $n_7=\binom{6}{2}=15$, and for $K=6$, $n_6=\binom{6}{3}=20$.
Each combination of zero and nonzero coefficients must be checked
with relations~\eqref{eq:A012345}. If the reconstruction of an
OS of the form~\eqref{eq:SS4} is impossible for some
combination of the coefficients, then this combination is
rejected. Such an analysis has been carried out in
Section~\ref{sec:3.2} and \ref{app:3} for $K=6$. As is indicated,
see Table~\ref{Tbl:2}, the number of possible OS's in
this case is $8$ which is much less than $n_6=20$.

A similar analysis has been done for $K=7$ and $K=8$. The results are
shown in Tables~\ref{Tbl:2} and~\ref{Tbl:3}. Hence, the total number
of possible OS's is $17$.

\subsection{Choosing an OS from
  candidate systems}\label{sec:3.4}

In order to choose an OS from candidate systems given in
Table~\ref{Tbl:2}, one can use an additional information about the
system. For example, if we know that the variable $x_1$ depends on
$x_2$ or $x_3$, then the OS must contain the coefficients
$a_2$ or $a_3$, correspondingly. If we know that the change of the
variable $x_3$ depends on its value, then the OS must
have the coefficient $c_3$, etc.
In other words, we need to find
features of the OS that single it out from other
candidate systems. Systems with no such features should be excluded
from the consideration.

There can arise situations where additional information regarding
the OS is not sufficient or is missing at all. The missing
features of the OS can be obtained in additional experiments if
the conditions of the experiment allow for a change of the
parameters, i.e., the coefficients of the OS. Namely, if the
conditions of the experiment change at least one of the
coefficients of the OS, then the observable $y_1(t)\equiv x_2(t)$
will change. The reconstruction in form (18) obtained from this
observable will have the value of at least one of the coefficients
$A_1, \dots, A_5$ changed. By analyzing these changes, one can
formulate new features of the sought OS.

\begin{figure}[ht]
\begin{center}
  \includegraphics[width=6cm]{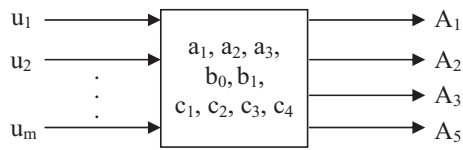}
\end{center}
  \caption{The scheme for determining the structure of the OS.}
  \label{fig:6}
\end{figure}

The study can be carried out along the following scheme, see
Fig.~\ref{fig:6}.

\begin{enumerate}

\item With some process conditions $(u_{11},...,u_{m1})$, we get a time series
  $y_{11}(t)$ for the observable $y_1(t)$ that is used for the
  reconstruction of type~\eqref{eq:SS4} with some coefficients
  $A_{11},A_{21}, A_{31}, A_{51}$.
\item For new process conditions $(u_{12}, \dots, u_{m2})$ we get a
  new time series $y_{12}(t)$ and new reconstruction of
  type~\eqref{eq:SS4} with new coefficients $A_{12}, A_{22}, A_{32},
  A_{52}$.
\item Repeat step~2 by trying to change other parameters if the
  experiment conditions permit to do so. For example, if step 2 has
  been performed with the process temperature changes, then we repeat
  it with the pressure changes, etc.
\item By performing step~3 with various process conditions, we
  determine which of the coefficients in the SS change.
\item Using the results of the experiments we formulate features for
  choosing the initial system.
\end{enumerate}

For making the analysis easier, one can use Table~\ref{Tbl:4}
obtained from Table~\ref{Tbl:3}. It clearly shows which of the
coefficients of various OS's determine the values of the
coefficients in the SS. For example, according to
Table~\ref{Tbl:4}, if only the coefficient $A_1$ is changed under
some actions in the SS~\eqref{eq:SS4}, this could have occurred
due to a change of $b_0$ or $b_1$ or $c_2$.  And if a coefficient
of the OS changes, this means that this coefficient is not zero.
Consequently, $b_0\neq 0$ or $b_1\neq 0$ or $c_2\neq 0$ in the OS.
Similarly, if only $A_2$ in the reconstructed system is changed
when different actions were applied, then this would mean that
$a_2\neq 0$ or $c_1\neq 0$ in the OS. If only $A_3$ in the SS is
changing, then one can assume that either $a_1\neq 0$ or $c_3\neq
0$ in the OS.

It is also convenient for making the analysis to use Table~\ref{Tbl:5}
obtained from Table~\ref{Tbl:4}. It shows which coefficients in the
SS may change when different actions are applied to the
OS. The information about the change of the coefficients
of the OS is not used in this case. When constructing
Table~\ref{Tbl:5}, various cases that may occur when external effects
are applied to the OS were considered. For example, if
the coefficient $A_1$ changes when using reconstruction~\eqref{eq:SS4}
from the observable in system $S_2$, then, according to
Table~\ref{Tbl:4}, this can occur if one of the coefficients of the
OS changes, namely the coefficient $c_2$. If system $S_1$
is considered, a change of $A_1$ can only occur if $b_1$ or $c_2$
change in the OS. All these cases are reflected in
Table~\ref{Tbl:5}.

A similar approach is used for combinations of $A$'s. For example, if
system $S_2$ is reconstructed, then simultaneous change of the
coefficients $A_1$ and $A_3$ is possible if the only coefficient $c_3$
in the OS is changed. But when reconstructing system
$S_7$, a change of $A_3$ can occur if $a_1$ is changed, whereas
changes $A_1$ if so does $c_2$. But if a change of the conditions of
the experiment implies simultaneous change of $a_1$ and $c_2$ in the
OS, that would imply that both $A_1$ and $A_3$ change.

Also, it is impossible for system $S_{12}$ that a change of only
one of the coefficients in the OS would cause simultaneous change
of the coefficients $A_1$, $A_2$, $A_5$ in the SS (and $A_3$
remains the same).  But a change of $a_2$ would imply changes of
$A_1$ and $A_2$, and a change of $a_3$ would imply the same for
$A_1$ and $A_5$.  If this occurs for the same external effect, we
get a number of changed coefficients $A_1$, $A_2$, $A_5$, which is
indicated in Table~\ref{Tbl:5}.

Combinations given in Table~\ref{Tbl:5} can be used as features
for choosing the OS. For example, if the SS shows a change of only
coefficient $A_2$, then, according to Table~\ref{Tbl:5}, this can
happen only if the OS's $S_1$, $S_3$, $S_5$, $S_7$, $S_9$,
$S_{11}$, $S_{13}$, $S_{14}$, $S_{15}$, $S_{16}$, and $S_{17}$ are
considered, and this can not happen for the OS's $S_2$, $S_4$,
$S_6$, $S_8$, $S_{10}$, $S_{12}$. An example of choosing the OS is
given in \ref{app:4}.

\newcommand{\eqlab}{$A_1,A_2$}
\newcommand{\eqlac}{$A_1,A_3$}
\newcommand{\eqlad}{$A_1,A_5$}
\newcommand{\eqlbc}{$A_2,A_3$}
\newcommand{\eqlbd}{$A_2,A_5$}
\newcommand{\eqlabc}{$A_1,A_2,A_3$}
\newcommand{\eqlabd}{$A_1,A_2,A_5$}

\begin{table}[ht]
  \centering
  \begin{tabular}{|c|c|c|c|c|c|c|c|c|c|}
    \hline
    OS &  \multicolumn{9}{|c|}{Coefficients of OS}\\
    \cline{2-10}
        &       $a_1$&  $a_2$&  $a_3$&      $b_0$&  $b_1$&  $c_1$&  $c_2$&  $c_3$&      $c_4$\\
    \hline

    $S_1$&      ---&    ---&    \eqlabd&    ---&    $A_1$&  $A_2$&  $A_1$&  $A_3$&      $A_5$\\
    \hline
    $S_2$&      ---&    \eqlab& \eqlad&     ---&    \eqlab& ---&    $A_1$&  \eqlac&     $A_5$\\
    \hline
    $S_3$&      ---&    \eqlab& \eqlbd&     ---&    \eqlab& $A_2$&  ---&    \eqlac&     $A_5$\\
    \hline
    $S_4$&      \eqlbc& ---&    \eqlad&     ---&    $A_1$&  ---&    $A_1$&  \eqlbc&     $A_5$\\
    \hline
    $S_5$&      $A_3$&  ---&    \eqlabd&    ---&    $A_1$&  $A_2$&  $A_1$&  ---&        $A_5$\\
    \hline
    $S_6$&      \eqlbc& \eqlab& $A_5$&      ---&    \eqlab& ---&    ---&    \eqlabc&    $A_5$\\
    \hline
    $S_7$&      $A_3$&  $A_2$&  \eqlad&     ---&    \eqlab& ---&    $A_1$&  ---&        $A_5$\\
    \hline
    $S_8$&      ---&    \eqlab& \eqlad&     $A_1$&  \eqlab& ---&    ---&    \eqlac&     $A_5$\\
    \hline
    $S_9$&      ---&    \eqlab& \eqlabd&    ---&    \eqlab& $A_2$&  $A_1$&  \eqlac&      $A_5$\\
    \hline

    $S_{10}$&   ---&    \eqlab& \eqlad&     $A_1$&  \eqlab& ---&    $A_1$&  \eqlac&     \eqlad\\
    \hline
    $S_{11}$&   \eqlbc& ---&    \eqlabd&    ---&    $A_1$&  $A_2$&  $A_1$&  \eqlbc&     $A_5$\\
    \hline
    $S_{12}$&   \eqlbc& \eqlab& \eqlad&     ---&    \eqlab& ---&    $A_1$&  \eqlabc&    $A_5$\\
    \hline
    $S_{13}$&   \eqlbc& \eqlab& \eqlbd&     ---&    \eqlab& $A_2$&  ---&    \eqlabc&    $A_5$\\
    \hline
    $S_{14}$&   $A_3$&  $A_2$&  \eqlabd&    ---&    \eqlab& $A_2$&  $A_1$&  ---&        $A_5$\\
    \hline
    $S_{15}$&   $A_3$&  $A_2$&  \eqlad&     $A_1$&  \eqlab& ---&    $A_1$&  ---&        \eqlad\\
    \hline
    $S_{16}$&   \eqlbc& \eqlab& \eqlabd&    ---&    \eqlab& $A_2$&  $A_1$&  \eqlabc&    $A_5$\\
    \hline
    $S_{17}$&   \eqlbc& \eqlab& \eqlabd&    $A_1$&  \eqlab& $A_2$&  $A_1$&  \eqlabc&    \eqlad\\
    \hline
  \end{tabular}
  \bigskip
  \caption{The connections between the coefficients of the SS's and
    the OS's according to Table~\ref{Tbl:3}.}
  \label{Tbl:4}
\end{table}

\begin{table}[ht]
  \centering
  \begin{tabular}{|c|c|c|c|c|c|c|c|c|c|c|c|c|c|c|c|c|c|}
    \hline
    Coefficient        &  \multicolumn{17}{|c|}{}\\
    combinations    &  \multicolumn{17}{|c|}{OS}\\
    \cline{2-18}
    for the SS's&   \small{$S_1$}&\small{$S_2$}& \small{$S_3$}& \small{$S_4$}& \small{$S_5$}& \small{$S_6$}& \small{$S_7$}&\small{$S_8$}&\small{$S_9$}
    &\small{$S_{10}$}&\small{$S_{11}$}&\small{$S_{12}$}&\small{$S_{13}$}&\small{$S_{14}$}&\small{$S_{15}$}& \small{$S_{16}$}& \small{$S_{17}$}\\
    \hline

    $A_1$&  \symb& \symb&      & \symb& \symb&      & \symb& \symb& \symb& \symb& \symb& \symb&      & \symb& \symb& \symb& \symb\\
    \hline
    $A_2$&  \symb&      & \symb&      & \symb&      & \symb&      & \symb&      & \symb&      & \symb& \symb& \symb& \symb& \symb\\
    \hline
    $A_3$&  \symb&      &      &      & \symb&      & \symb&      &      &      &      &      &      & \symb& \symb&      &      \\
    \hline
    $A_5$&  \symb& \symb& \symb& \symb& \symb& \symb& \symb& \symb& \symb&      & \symb& \symb& \symb& \symb&      & \symb&      \\
    \hline
    \eqlab& \symb& \symb& \symb&      & \symb& \symb& \symb& \symb& \symb& \symb& \symb& \symb& \symb& \symb& \symb& \symb& \symb\\
    \hline
    \eqlac& \symb& \symb& \symb&      & \symb&      & \symb& \symb& \symb& \symb&      &      &      & \symb& \symb&      &      \\
    \hline
    \eqlad& \symb& \symb&      & \symb& \symb&      & \symb& \symb& \symb& \symb& \symb& \symb&      & \symb& \symb& \symb& \symb\\
    \hline
    \eqlbc& \symb&      &      & \symb& \symb& \symb& \symb&      &      &      & \symb& \symb& \symb& \symb& \symb& \symb& \symb\\
    \hline
    \eqlbd& \symb&      & \symb&      & \symb&      & \symb&      & \symb&      & \symb&      & \symb& \symb&      & \symb&      \\
    \hline
    $A_3,A_5$&  \symb&      &      &      & \symb&      & \symb&      &      &      &      &      &      & \symb&      &      &      \\
    \hline
    \eqlabc&    \symb& \symb& \symb& \symb& \symb& \symb& \symb& \symb& \symb& \symb& \symb& \symb& \symb& \symb& \symb& \symb& \symb\\
    \hline
    \eqlabd&    \symb& \symb& \symb&      & \symb& \symb& \symb& \symb& \symb& \symb& \symb& \symb& \symb& \symb& \symb& \symb& \symb\\
    \hline
 $A_1,A_3,A_5$& \symb& \symb& \symb&      & \symb&      & \symb& \symb& \symb& \symb&      &      &      & \symb& \symb&      &      \\
    \hline
 $A_2,A_3,A_5$& \symb&      &      & \symb& \symb& \symb& \symb&      &      &      & \symb& \symb& \symb& \symb&      & \symb&      \\
    \hline
$A_1,A_2,A_3,A_5$&  \symb& \symb& \symb& \symb& \symb& \symb& \symb& \symb& \symb& \symb& \symb& \symb& \symb& \symb& \symb& \symb& \symb\\
    \hline
  \end{tabular}
  \bigskip
  \caption{Combinations of coefficients of the SS, which can change while the OS undergoes external effects.}
  \label{Tbl:5}
\end{table}

\section{Conclusions}\label{sec:4}

The problem of reconstructing an original system, in general, is
very complex and is solved in present primarily using numerical
methods. At the same time, an analytic approach may be effective
in particular cases. This paper shows a relative rigor of this
approach if applied to SR-class systems. The method can also be
applied to other classes of systems if there are transformations
that connect the standard and the original systems.

This approach allows not only to obtain new systems that make an
alternative to the known ones, as shown with an example of the
\Rossler system, but also to establish a set of candidate systems
used to choose a unique real original system. Naturally, to solve
the latter problem, a use of one observable is not sufficient. One
also needs additional information about the system or a
possibility to interact with the system in order to obtain such
information. If the obtained information permits, we can determine
not only the structure of the sought original system but also
relations between its coefficients.

\appendix
\section{}\label{app:1}

To prove Proposition~\ref{P:1}, let us find a relation between the
variables in systems~(\ref{eq:10}) and~(\ref{eq:11}).

Since $y_1=x_2$, the second equation in~(\ref{eq:10}) can be written as
\begin{equation}
  \label{eq:y1t}
  \dot y_1=b_0+b_1 x_1 + b_2 y_1,
\end{equation}
whence,
\begin{equation}
  \label{eq:x1}
  x_1=\frac{\dot y_1 -b_2 y_1 -b_0}{b_1}.
\end{equation}
By differentiating~(\ref{eq:y1t}) with respect to $t$ we get
\begin{equation}
  \label{eq:y12t}
  \ddot y_1= b_1 \dot x_1+ b_2 \dot y_1.
\end{equation}
Substituting the expressions for $\dot x_1$ and $\dot y_1=\dot x_2$
from~(\ref{eq:10}) into~(\ref{eq:y12t}) and making simplifications we get
\begin{equation}
  \label{eq:y12t1}
  \ddot y_1=a_0 b_1 + b_0 b_2 +
  (a_1 b_1 + b_1 b_2) x_1 + (a_2 b_1 + b_2^2) x_2 + a_3 b_1 x_3.
\end{equation}
Substitute now $x_1$ from~(\ref{eq:x1}) into~(\ref{eq:y12t1}) to find
\begin{equation}
  \label{eq:x3}
  x_3 =
  \frac{\ddot y_1 - ( a_1 + b_2 )\dot y_1 + ( a_1b_2 - a_2b_1 ) y_1 +
    a_1b_0 - a_0b_1}
  {a_3b_1}.
\end{equation}
Differentiating~(\ref{eq:y12t1}) with respect to $t$ we get
\begin{equation*}
    \dddot y_1 = ( a_1b_1 + b_1b_2 ) \dot x_1 +
    ( a_2b_1 + b_2^2 ) \dot x_2 + a_3b_1 \dot x_3.
\end{equation*}
After making necessary substitutions from~(\ref{eq:10}) we find
\begin{equation}
  \label{eq:y13t}
  \begin{aligned}
    \dddot y_1 =& ( a_1b_1 + b_1b_2 )
    ( a_0 + a_1 x_1 + a_2 x_2 + a_3 x_3 ) +
    ( a_2b_1 + b_2 ^2) \dot x_2 \\
    &+ a_3b_1 ( c_0 + c_1 x_1 + c_2 x_2 +
    c_3 x_3 + c_4 x_1 x_2 + c_5 x_1 x_3 + c_6 x_2 x_3 ) .
  \end{aligned}
\end{equation}
Replace $x_1$ and $x_3$ in~(\ref{eq:y13t}) using~(\ref{eq:x1})
and~(\ref{eq:x3}), correspondingly, and $\dot x_2$ with $\dot
y_1$. By grouping similar terms, we get
\begin{equation*}
  \dddot y_1 = A_0 + A_1  y_1 + A_2 \dot y_1 +
  A_3 \ddot y_1 + A_4 y_1^2 + A_5 y_1\dot  y_1 +
  A_6 y_1 \ddot y_1 + A_7\dot y_1^2 + A_8 \dot y_1\ddot y_1,
\end{equation*}
where the coefficients $A_0,\dots,A_8$ are given in
system~(\ref{eq:12}). Setting $\dot y_1=y_2$ and $\ddot y_1 = y_3$ we
get system~(\ref{eq:11}). The proposition is proved.

\section{}\label{app:2}

Comparing relations~(\ref{eq:14}) and~(\ref{eq:16}) we find that
the expressions for $A_0 $, $A_4 $, $A_5 $, $A_6 $, $A_7 $, $A_8 $
in these cases coincide and those for $A_1 , A_2 , A_3 $ differ.
To find $b_0$ in system~(\ref{eq:16}), we set $a_2^{13} = a_2^{15}
= a_2 $, $a_3^{13} = a_3^{15} = a_3 $, $b_1^{13} = b_1^{15} = b_1
$, $b_2^{13} = b_2^{15} = b_2 $, $c_0^{13} = c_0^{15} = c_0$,
$c_5^{13} = c_5^{15} = c_5 $, where the upper indices indicate
whether the coefficients enter system~(\ref{eq:13})
or~(\ref{eq:15}), respectively. The new coefficient $b_0^{15}$ is
found, for example, by equating the relations for $A_3$
in~(\ref{eq:14}) and~(\ref{eq:16}). Then $b_2^{13}+c_3^{13} =
b_2^{15} - b_0^{15} c_5^{15}/b_1^{15}$. Since $b_2^{13}=b_2^{15}$,
we have that $b_0^{15} =-c_3^{13} b_1/c_5$. It follows
from~(\ref{eq:13}) that the numerical value of the new coefficient
will be $b_0^{15}=c$. Substituting the relation for $b_0^{15}$
into the expression for $A_1$ and $A_2$ from~(\ref{eq:16}) we get
$A_1=-a_2 b_1 c_3^{13}$ and $A_2=a_2 b_1- b_2 c_3^{13}$, that are
similar to the expressions for $A_1$ and $A_2$ in~(\ref{eq:14}).
Consequently, numerical values of all the coefficients in the
reconstructions of systems~(\ref{eq:13}) and~(\ref{eq:15})
coincide and, hence, system~(\ref{eq:16}) meets the conditions in
Proposition~\ref{P:2}.

\section{}\label{app:3}

Let us analyze expression~(\ref{eq:12}) by comparing~(\ref{eq:11}) with~(\ref{eq:SS4}). Since $A_8=0$ in~(\ref{eq:SS4}), we have
\begin{equation}
  \label{eq:c5}
  c_5=0
\end{equation}
in the OS by~(\ref{eq:12}). Consequently, $A_7=0$, which
is verified by the reconstruction~(\ref{eq:SS4}). It also follows
from~(\ref{eq:c5}) that $A_6=c_6$. But $A_6=0$ in~(\ref{eq:SS4}) and,
consequently,
\begin{equation}
  \label{eq:c6}
  c_6=0.
\end{equation}
Since $A_5\neq 0$ but~(\ref{eq:c5}) and~(\ref{eq:c6}) hold, we have
\begin{equation}
  \label{eq:a3c4}
  a_3\neq 0,\qquad c_4\neq 0.
\end{equation}
Because $A_4=0$, it follows from~(\ref{eq:c5}),~(\ref{eq:c6}),
and~(\ref{eq:a3c4}) that
\begin{equation}
  \label{eq:b2}
  b_2=0.
\end{equation}
Consequently,
\begin{equation}
  \label{eq:b1}
  b_1\neq 0,
\end{equation}
for, otherwise, the second equation in~(\ref{eq:10}) splits from the
other equations.

In view of~(\ref{eq:c5}),~(\ref{eq:c6}), and~(\ref{eq:b2}),
system~(\ref{eq:10}) becomes
\begin{equation*}
  \left\{
    \begin{aligned}
      \dot x_1 &= a_0 + a_1 x_1 + a_2 x_2 + a_3 x_3 ,\\
      \dot x_2 &= b_0 + b_1 x_1 ,\\
      \dot x_3 &= c_0 + c_1 x_1 + c_2 x_2 + c_3 x_3 + c_4 x_1 x_2 .
    \end{aligned}
  \right.
\end{equation*}
Let us find the coordinates $x_{10}, x_{20}, x_{30}$ of the
equilibrium position of the system by setting $\dot x_1, \dot x_2,
\dot x_3$ to zero,
\begin{equation}
  \label{eq:SR0}
  \left\{
    \begin{aligned}
      a_0 + a_1 x_{10} + a_2 x_{20} + a_3 x_{30} &= 0,\\
      b_0 + b_1 x_{10} &= 0,\\
      c_0 + c_1 x_{10} + c_2 x_{20} + c_3 x_{30} + c_4 x_{10} x_{20}
      &= 0.
    \end{aligned}
  \right.
\end{equation}
Since the observable $y_1(t)=x_2(t)$ oscillates about zero, see
Fig.~\ref{fig:4}, we have $x_{20}=0$ in~(\ref{eq:SR0}). Then the
system becomes
\begin{equation}
  \label{eq:SR01}
  \left\{
    \begin{aligned}
      a_0 + a_1 x_{10}  + a_3 x_{30} &= 0,\\
      b_0 + b_1 x_{10} &= 0,\\
      c_0 + c_1 x_{10}  + c_3 x_{30}
      &= 0.
    \end{aligned}
  \right.
\end{equation}
The second equation in~(\ref{eq:SR01}) gives
\begin{equation}
  \label{eq:x10}
  x_{10}=-\frac{b_0}{b_1}.
\end{equation}
Substituting~(\ref{eq:x10}) into the first and the third equations
in~(\ref{eq:SR01}) we get
\begin{equation}
  \label{eq:SR02}
  \left\{
  \begin{aligned}
    x_{30}&=
     \frac{a_1b_0 / b_1 - a_0 }{  a_3 },\\
    x_{30} &= \frac{c_1b_0 / b_1 - c_0 } { c_3} .
  \end{aligned}
\right.
\end{equation}
By equating the right-hand sides of~(\ref{eq:SR02}), we have
\begin{equation}
  \label{eq:SR03}
  \frac{a_1b_0 - a_0b_1 }{ a_3} = \frac{c_1b_0 - c_0b_1 }{ c_3} .
\end{equation}
We have assumed, see Section~\ref{sec:3.4}, that the values of the
coefficients of the OS may change in an arbitrary way if
the system undergoes different effects, but relation~(\ref{eq:SR03})
must always hold. This permits us to make the following conclusions.
\begin{enumerate}
\item The expressions in the left- and the right-hand sides
  of~(\ref{eq:SR03}) must be equal to zero. Otherwise, the relation
  could change if the coefficients entering this relation
  change. Consequently,
  \begin{equation}
    \label{eq:SR04}
    a_1b_0 - a_0b_1 = c_1b_0 - c_0b_1 = 0.
  \end{equation}
\item Moreover,~(\ref{eq:SR03}) could fail to hold if the coefficients
  of the OS change in the case where some of the four
  terms in~(\ref{eq:SR03}) or~(\ref{eq:SR04}) are distinct from
  zero. Hence, the following must be true:
  \begin{equation}
    \label{eq:SR05}
    a_1b_0 = a_0b_1 = c_1b_0 = c_0b_1 = 0.
  \end{equation}
\end{enumerate}
Note that $A_0=0$ in system~(\ref{eq:A012345}) as follows
from~(\ref{eq:SR05}), which is needed for the SS to
be of the form~(\ref{eq:SS4}).

For~(\ref{eq:SR05}) (and~(\ref{eq:SR04})) to hold, it is
sufficient that each of the monomials would have at least one
coefficient zero. By~(\ref{eq:b1}), this implies that
\begin{equation}
  \label{eq:a0c0}
  a_0=c_0=0.
\end{equation}
Moreover, at least one of the two following conditions must be satisfied:
\begin{equation}
  \label{eq:b0}
  b_0=0
\end{equation}
or
\begin{equation}
  \label{eq:a1c1}
  a_1=c_1=0.
\end{equation}
So, by now we have that all candidate systems must have $a_0 = b_2
= c_0 = c_5 = c_6 = 0$ and $ a_3\neq 0$, $b_1\neq 0$, $c_4\neq 0
$. The other coefficients that enter the right-hand sides of
equations~(\ref{eq:A012345}), $a_1 , a_2 , b_0 , c_1 , c_2 , c_3 $
may or may not be equal to zero in different versions of the
OS.

Using~(\ref{eq:SR04}) we get from~(\ref{eq:SR02}) and~(\ref{eq:SR03})
that
\begin{equation}
  \label{eq:x30}
  x_{30}=0.
\end{equation}
Substituting~(\ref{eq:a0c0}) and~(\ref{eq:x30}) into~(\ref{eq:SR01})
we get
\begin{equation}
  \label{eq:abc}
  \left\{
  \begin{aligned}
     a_1 x_{10}& = 0,\\
     b_0 + b_1 x_{10} &= 0,\\
     c_1 x_{10} &= 0.
   \end{aligned}
 \right.
\end{equation}
If~(\ref{eq:b0}) holds, since $b_1\neq0$, the second equation
in~(\ref{eq:abc}) gives that $x_{10}=0$. If~(\ref{eq:a1c1}) holds
and $b_0\neq 0$, then $x_{10}=-b_0/b_1$.

Let us analyze the cases of~(\ref{eq:b0}) and~(\ref{eq:a1c1}).
\begin{enumerate}
\item If $b_0=0$, then $a_0 = b_0 = b_2 = c_0 = c_5 = c_6 = 0 $ and
  the candidate systems can contain $8$ nonzero coefficients $a_1 ,
  a_2 , a_3 , b_1 , c_1 , c_2 , c_3 , c_4 $ where $a_3 , b_1 , c_4 $
  are always nonzero. Since the candidate systems have $6$ nonzero
  coefficients in the right-hand sides, all possible combinations of
  $3$ more coefficients out of the remaining $5$ coefficients $a_1 ,
  a_2 , c_1 , c_2 , c_3 $ should be considered in view of
  Proposition~2. If two of them are zero, then the remaining three
  coefficients, together with $a_3 , b_1 , c_4$, will determine the
  structure of the candidate system. If we substitute the zero
  coefficients into system~\eqref{eq:A012345}, we obtain one of the
  systems $S_1$ -- $S_7$ given in Table~\ref{Tbl:2}. A system with
  $a_1=a_3=0$ has been excluded from possible OS's, since
  $A_3=0$ in this case, which contradicts the
  SS~(\ref{eq:SS4}). Systems with $a_2=c_2=0$ or with $c_2=c_3=0$
  should also be disregarded, since that would give $A_1=0$.
\item If $b_0\neq 0$ and $a_1=c_1=0$, we get $a_0 = a_1 = b_2 = c_0 =
  c_1 = c_5 = c_6 = 0 $, $a_3 \neq 0, b_0 \neq 0, b_1 \neq 0 , c_4
  \neq 0 $. Since the OS must have $6$ nonzero
  coefficients, the remaining two must be chosen from $a_2 , c_2 , c_3
  $. Set one of them to be zero each time and enter the others into
  the OS. It turns out that for $a_2=0$, $A_2=0$
  by~(\ref{eq:A012345}), and if $c_3=0$, then $A_3=0$, which contradicts
  the form of the SS~(\ref{eq:SS4}). If $c_2=0$, the
  structure of the SS~(\ref{eq:SS4}) does not change, so a
  system with nonzero $a_2 , a_3 , b_0 , b_1 , c_3 , c_4 $
  (system~$S_8$ in the Table~\ref{Tbl:2}) can be regarded as a candidate system.
\end{enumerate}

\section{}\label{app:4}

Let us consider an example of reconstruction of an SR-class OS
from one observable in the case where the SS is of
form~\eqref{eq:SS4}.  It follows from the analysis carried out in
\ref{app:3} that the only nonzero coefficients the OS
can have are $a_1$, $a_2$, $a_3$, $b_0$, $b_1$, $c_1$, $c_2$,
$c_3$, $c_4$ satisfying relations~\eqref{eq:A012345}. Its general form
is given in~\eqref{eq:k9}, and all possible choices for the OS
are given in Table~\ref{Tbl:2}.

Assume we know that a change of the variable $x_1$ in the sought
system does not depend on the value of the variable $x_1$.  In other
words, the coefficient $a_1$ in system (24) is zero. We will consider
this fact as the first feature for choosing the OS from
all candidate systems. As Table~\ref{Tbl:4} shows, systems $S_1$,
$S_2$, $S_3$, $S_8$, $S_9$, $S_{10}$ have this property.

Assume that measurements of the observable were taken varying the
conditions of the experiment and value of the only coefficient $A_2$
was changed in reconstruction~\eqref{eq:SS4}, while the other
coefficients $A_1$, $A_3$, $A_5$ remained unchanged. We then can use
this as a second feature to single out the sought OS from
other candidates. Table~\ref{Tbl:5} shows that the second feature is
enjoyed by the systems $S_1$, $S_3$, $S_5$, $S_7$, $S_9$, $S_{11}$,
$S_{13}$, $S_{14}$, $S_{15}$, $S_{16}$, $S_{17}$. To make the analysis
easier, we summarize this information in Table~\ref{Tbl:6}.

\begin{table}[ht]
  \centering
  \begin{tabular}{|c|c|c|c|c|c|c|c|c|c|c|c|c|c|c|c|c|c|c|}
    \hline
    \multicolumn{2}{|c|}{OS}&
        $S_1$& $S_2$& $S_3$& $S_4$& $S_5$& $S_6$& $S_7$& $S_8$& $S_9$&$S_{10}$&$S_{11}$&$S_{12}$&$S_{13}$&$S_{14}$&
    $S_{15}$&$S_{16}$&$S_{17}$\\
    \hline
    Features
&   1&  \symb& \symb& \symb&      &      &      &      & \symb& \symb& \symb&      &      &      &      &      &      &      \\
    \cline{2-19}
&   2&  \symb&      & \symb&      & \symb&      & \symb&      & \symb&      & \symb&      & \symb& \symb& \symb& \symb& \symb\\
    \hline
  \end{tabular}
  \bigskip
  \caption{Distribution of features 1 and 2 in candidate systems.}
  \label{Tbl:6}
\end{table}

As the table shows, only systems $S_1$, $S_3$, and $S_9$ possess
features 1 and 2. Hence, the number of systems to consider is
reduced from 17 to 3. System $S_1$ is excluded from the
considerations, since it has the same structure as the SS, for
otherwise the problem could be considered as solved after
obtaining the reconstruction~\eqref{eq:SS4}. Systems $S_3$ and
$S_9$ have, respectively, 6 and 7 nonzero coefficients in the
right-hand side, with the coefficients $a_2$, $a_3$, $b_1$, $c_1$,
$c_3$, and $c_4$ entering both systems, whereas the coefficient
$c_2$ is nonzero only in system $S_9$. If the information at hand
does not allow to uniquely choose one of the systems, then it
seems reasonable to choose a simpler system, which is $S_3$ in
this case. Let us remark that this system, which is the same as
system~\eqref{eq:OS6}, was considered as an example of
reconstruction in Section~\ref{sec:3.2}.

\bibliographystyle{model1-num-names}
\bibliography{gorodetskyi}

\end{document}